\documentclass[journal=jacsat,manuscript=article]{achemso}
\usepackage{graphicx} 
\usepackage{amsmath}
\usepackage{amssymb}

\renewcommand{\Im}{\operatorname{Im}}
\usepackage{url}
\usepackage[version=3]{mhchem} 

\title{Inverse design of 3D-printable metalenses with complementary dispersion for terahertz imaging}

\author{Mo Chen}
\affiliation{Department of Mathematics, Massachusetts Institute of Technology, Cambridge, Massachusetts 02139, USA}
\author{Ka Fai Chan}
\affiliation{State Key Laboratory of Terahertz and Millimeter Waves, CityU, Hong Kong, China}
\author{Alec M. Hammond}
\affiliation{School of Electrical and Computer Engineering, Georgia Institute of Technology, Atlanta, GA 30308, USA}
\author{Chi Hou Chan}
\affiliation{State Key Laboratory of Terahertz and Millimeter Waves, CityU, Hong Kong, China}
\author{Steven~G.~Johnson}
\affiliation{Department of Mathematics, Massachusetts Institute of Technology, Cambridge, Massachusetts 02139, USA}
\email{stevenj@math.mit.edu}

\begin{document}

\maketitle

\begin{abstract}
    This study formulates a volumetric inverse-design methodology to generate a pair of complementary focusing metalenses for terahertz imaging: the two lenses exhibit equal and opposite shifts in focal length with frequency.  An asymmetry arises, where we find a focal length that decreases with frequency to be more challenging to achieve (without material dispersion) given fabrication constraints, but it is still possible.   We employ topology optimization, coupled with manufacturing constraints, to explore fully freeform designs compatible with 3D printing.   Formulating an optimization problem that quantifies the goal of maximal complementary focal shifts, while remaining differentiable and tractable, requires a carefully selected sequence of constraints and approximations.
\end{abstract}

\section{Introduction}

\begin{figure}[ht!]
    \centering
    \includegraphics[width=0.9\textwidth]{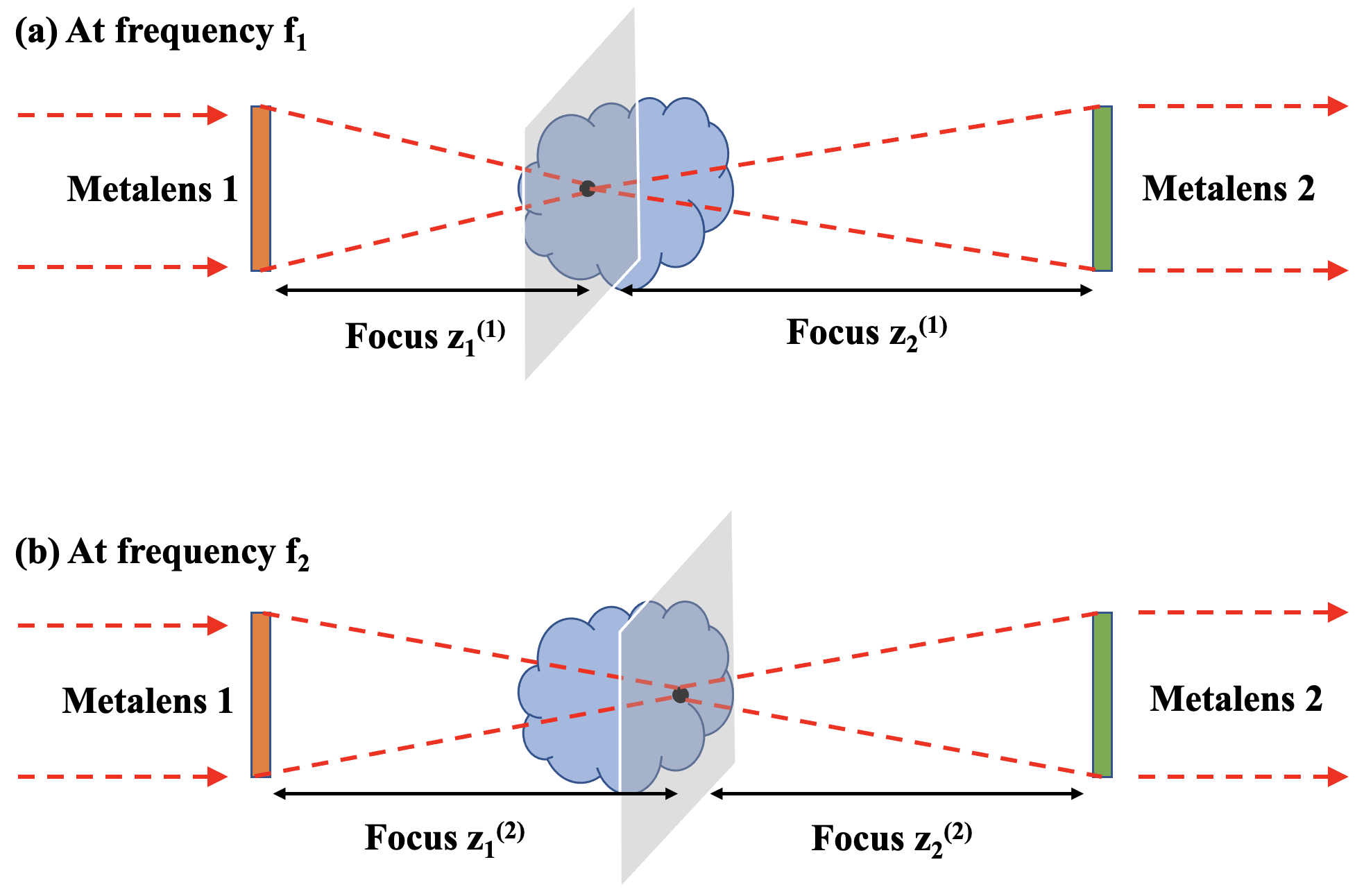}
    \caption{A pair of metalenses with complementary scanning capabilities: the two foci move in equal and opposite directions as frequency changes, allowing scanning of internal structure.}
    \label{fig:setup}
\end{figure}

In this work, we employ freeform topology optimization to design a pair of 3D metalenses with complementary focus-scanning capability:  one metalens exhibits an \emph{increase} in focal length with increasing frequency  (``diffractive'' behavior~\cite{Stone88}, analogous to Fresnel lenses~\cite{Engelberg2020,Khorasaninejad17}), while the other shows an equal and opposite \emph{decrease} in focal length (``refractive'' behavior~\cite{Stone88}, analogous to the effect of ``normal''~\cite{born2013principles} material dispersion in refractive lenses). As illustrated in Fig.~\ref{fig:setup}, such a pair of metalenses could collectively form an imaging system capable of scanning the internal structure of objects placed between them. A traditional Fresnel lens also has a focal length that increases with frequency\cite{PFL1961}, and correspondingly we find that obtaining this behavior is a more straightforward design problem for metalenses. In contrast, we find that the other ``refractive'' direction is more challenging for a thin lens with negligible material dispersion (compared to the desired 15--20\% focal shift), but remains possible (because other forms of dispersion can be introduced geometrically, e.g.~by resonances), and the focal quality is more limited by the imposed fabrication constraints. In both cases, we employed freeform 3D topology optimization with lengthscale and connectivity constraints to design axisymmetric structures that are compatible with 3D printing in the terahertz regime; in addition, we introduced a damping term in the formulation to facilitate the binarization of structure into solid or air. The entire optimization was divided into three stages: first, we optimized for a monochrome focus to determine the attainable intensity; second, we maximized an efficient estimate of the rate of change of focal length with frequency, subject to a constraint on focal intensity relative to the monochrome focus; third, we optimized the frequency-shifted focal quality for a rate of focal shift determined by the second stage. As a consequence, our metalenses have both good focal qualities and large complementary focal-length changes ($30.4 \to 35.6$~mm and $52.2 \to 47$~mm from 250 GHz to 300 GHz).  The empirical asymmetry of the design problem for opposite signs of the dispersion also invites future study on the fundamental limitations of such processes~\cite{Chao2022}.

The proposed metalenses are designed to operate at terahertz (THz) frequencies. Terahertz waves possess various appealing features, such as wide bandwidth, non-ionization, and penetration capabilities, and are rapidly emerging as the new frontier of electromagnetic research\cite{Siegel02,Song11}. THz imaging has a higher resolution than the microwave due to the shorter wavelength, can penetrate obscuring materials such as clothing, and is significantly safer for humans than X-rays~\cite{Chan20,Chan22}. A key area of interest in THz imaging is focus-scanning capability, which has been explored using various techniques~\cite{Chan20,liu23}. One idea involves manipulating focal points simply by varying the incident frequencies within a specified range using reflective metalenses~\cite{Yi2017FlatTR}, and that work directly inspired us to design an analogous transmissive system (Fig.~1): a pair of compact (few-wavelengths-thick) with complementary focus--frequency dependence.  Metalenses are an attractive candidate for this design problem: they employ subwavelength structures that exploit the fullwave scattering physics, potentially offering the greatest design freedom for ultra-thin focusing, and they have been used to achieve a variety of unconventional lens-like effects~\cite{yu2014flat,Beneck24,Khorasaninejad17,chen2018broadband,Cui_2024,Kuznetsov24,wang23,roberts20233d}.

The greatest design freedom is achieved by solving the full Maxwell equations (as opposed to localized approximations~\cite{Pestourie18,Lin19,LPA2019}), which is practical for small-diameter (tens of wavelengths) THz lenses, especially for axisymmetric designs~\cite{Christiansen20,lin2021computational}.  Fabrication by 3D printing offers a huge space of design parameters, which can be explored computationally by large optimization.  In particular, we employ volumetric density-based topology optimization~\cite{Maute13}, in which every ``voxel'' of the design is allowed to vary smoothly between solid and void, with additional constraints~\cite{Hammond21DRC,Christiansen_2023} to ensure that the final design converges to a manufacturable ``binary'' solid--void structure.  Topology optimization has been applied to a variety of problems in optics~\cite{jensen2011topology}, including the design of axisymmetric metalenses~\cite{Christiansen20, lin2021computational}. For scanning metalenses, a part of the challenge we address in this paper is to mathematically formulate a tractable optimization problem that balances the tradeoffs of focal quality (efficiency and spot size) and scanning rate (frequency dependence).

\section{Physics Model and Optimization}
\subsection{Topology Optimization in Cylindrical Coordinates} \label{sec:to_review}

We will first review the basic ideas of topology optimization\cite{Maute13}. In this case, we are only interested in the field intensities along the central axis. Due to such axisymmetry of the objective, we will use cylindrical coordinates and express Maxwell's equations for the electric field $\textbf{E}(r,\phi,z)$ in the frequency domain at angular frequency $\omega$ as follows:
 \begin{equation}\label{eq:Maxwell}
     \left[\nabla \times \nabla \times {} - \frac{\omega^2}{c^2}\varepsilon(r,\phi,z)\right]\textbf{E}(r,\phi,z) = i\omega\mu_0 \textbf{J}(r,\phi,z), \quad (r,\phi,z) \in \Omega
 \end{equation}
where $c$ is the speed of light in vacuum, $\mu_0$ is the vacuum magnetic permeability, $\varepsilon(r,\phi, z)$ is the relative permittivity in space (omitting $\omega$ dependence: we assume negligible material dispersion in this paper), and $\Omega \subset \mathbb{R}^3 $ is the modelling domain. Additionally, we use perfectly matched layers (PML) \cite{BEREGNER_1994,Sacks1995PML,OSKOOI2011PML} to impose outgoing/radiation boundary conditions.   Although this equation is expressed in the frequency domain (time-harmonic fields $\sim e^{-i\omega t}$), we employ a hybrid time--frequency topology-optimization algorithm~\cite{Hammond2022} that allows us to optimize over a broad bandwidth using a single time-domain simulation, implemented in a free/open-source finite-difference time-domain (FDTD) software package~\cite{oskooi2010meep}.

To reduce the computational cost, one can restrict the degrees of freedom to axisymmetric structures~\cite{Christiansen20,lin2021computational}. Thus, we impose cylindrical symmetry, such that $\varepsilon(r,\phi,z) = \varepsilon(r,z)$, and decompose the electric field $\textbf{E}(r,\phi,z)$ as $\textbf{E}(r,\phi,z) = \sum_{m \in \mathbb{Z}} \textbf{E}_m(r,z)e^{im\phi}$. A circularly polarized normal-incident planewave corresponds to $m = \pm 1$, and it is sufficient to optimize the lens for $m=1$ (a mirror flip yields $m=-1$, and superposition yields any desired polarization).

In density-based topology optimization, one parameterizes $\varepsilon(r,z)$ within a design region $\Omega_d \subset \Omega $ using raw design variables $\rho(r,z)$ that varies continuously between~0 and~1. A value of $\rho(r,z)=0$  corresponds to the absence of material (air) at that point, while $\rho(r,z)=1$ corresponds to solid material (dielectric) at that point. In practice, to avoid arbitrarily small design features and to ensure convergence with resolution~\cite{sigmund1998numerical,sigmund2007morphology}, one applies an additional smoothing step to~$\rho(r,z)$:
\begin{equation}
    \Tilde{\rho}(r,z) = w(r) * \rho(r,z)
\end{equation}
where $*$ is convolution and $w$ is a conic filter kernel.  This $\Tilde{\rho}$ is then passed through a smoothed/differentiable ``step'' function, with a steepness parameter $\beta$, to obtain a nearly binarized density $\hat{\rho}$~\cite{Wang2011, lazarov2016length}:
\begin{equation}\label{eq:tanh_proj}
    \hat{\rho}(r,z) = \frac{\tanh(\beta/2) + \tanh(\beta (\Tilde{\rho}(r,z) -0.5))}{2\tanh (\beta/2)} \, .
\end{equation}
The hyper-parameter $\beta \in [1, \infty]$ is gradually increased during optimization to ensure that $\hat{\rho}(r,z)$ converges to binary values of~0 or~1 almost everywhere in $\Omega_d$. Finally, one obtains the physical dielectric constant $\varepsilon(r,z)$ by linearly interpolating $\hat{\rho}(r,z)$:
\begin{equation}
    \varepsilon(r,z) = 1 + \hat{\rho}(r,z)(\varepsilon_{\text{max}} - 1) \, 
\end{equation}
where we use $\varepsilon_{\text{max}} = 2.66$ corresponding to the resin used in the 3D printing approach demonstrated in Ref~\citenum{Wu2019} at a frequency of $300$~GHz~\cite{Chan20}.

Since we are interested in the field values far from the structure, one can reduce the computational cost by only collecting the fields directly above the structure and applying a near-to-far--field transformation~\cite{Taflove2005,oskooi2013}. This method allows one to efficiently compute the field at any distant points without actually including those points in the simulation cell. The near-to-far--field transformation is essentially a convolution with the Green's function, and its adjoint gradient can be efficiently computed by applying an additional chain rule through the transposed Green's-function convolution~\cite{Hammond2022}.

For a general inverse-design problem, the goal is to find the structure ($\rho$) that maximizes some real-valued figure-of-merit $\Phi(\textbf{E}(r,z))$. 
Large-scale optimization algorithms require the gradient $\nabla_\rho \Phi$, which can be efficiently computed (requiring only a single additional Maxwell solve) by  adjoint sensitivity analysis~\cite{jensen2011topology, niederberger2014sensitivity}.  Specifically, we used CCSA~\cite{Svanberg2002} optimization algorithm in this paper, implemented in a free software package~\cite{nlopt} . The choice of the precise figure of merit for a given engineering goal is often non-obvious, and we discuss our specific formulation for the metalens-scanning problem in the next section.

To ensure minimum-lengthscale for both the solid and void regions, we imposed geometric constraints described in Ref.~\citenum{Zhou15}, which were previously implemented in Ref.~\citenum{Hammond21DRC}; to ensure that all solid materials are connected to the substrate and that the void is connected to the air above, we implemented and imposed connectivity constraints described in Ref.~\citenum{Christiansen_2023} and Ref.~\citenum{Li2016}.
Projection by Eq.~\ref{eq:tanh_proj} may not be sufficient to ensure binarization if the physics strongly favors intermediate materials---in this case, the density  $\rho$ may converge to values around $0.5$, which will be projected to intermediate values of $\hat{\rho}$. This turns out to be the case for our metalens design problem, which  ideally might employ continuously varying materials to ``adiabatically'' maximize transmission~\cite{Christiansen_2023}.  To combat this tendency, we introduced an artificial damping term $\omega \Im \varepsilon \sim \hat{\rho} (1 - \hat{\rho})$ that further penalizes intermediate values~\cite{jensen2005damping, hansen2024inverse}.

\subsection{Optimization of Metalens with Scanning Capability}

As described in the introduction section, the entire optimization has three stages. Ultimately, we want to obtain equal yet large focal shifts in opposite directions, so we first need to find a reasonable shift magnitude as the target. Identifying the maximal achievable shift via optimization is nontrivial because obtaining focal peak itself requires a separate inner optimization. Instead, we formulated an estimate that efficiently evaluates the focal locations and the shifts without nested optimization. However, in maximizing the focal shift, we do not wish to sacrifice too much focal intensity compared to a simple lens, so we also perform a preliminary monochromatic-lens optimization to obtain a reference value that constrains the subsequent intensities.

\subsubsection{Stage one: Obtaining reference focal intensity}
To begin with, we simply design a monochromatic lens: we maximize the field intensity~\cite{Pestourie18} at a focal length $z_0$ for frequency $f_0$, without any fabrication constraints, and obtain the reference intensity $I_0$.  Here, $z_0$ is chosen to be 35 mm (based on a typical laboratory setup) and $f_0$ is chosen to be 275~GHz (in the middle of the bandwidth).   In subsequent steps, we  constrain the focal intensity to be $\ge 0.9 I_0$.

\subsubsection{Stage two: Determining maximal focal shift}\label{sec:max_shift}
We would like to estimate the maximally attainable frequency-dependent focal shift around $f_0$ and $z_0$: 
\begin{equation}\label{Eq:max_shift}
    \max_\rho \quad \pm \left[ z(\rho, f_0+\delta f)-z(\rho, f_0-\delta f) \right]
\end{equation}
where $\delta f$ is a given frequency shift and $z(\rho, f)$ are the focal positions at frequency $f$, respectively.  (Equivalently, for small $\delta f$, we would like to maximize $\pm \partial z / \partial f$.)  However, determining $z(\rho)$ would require a nested optimization (for a given $\rho$) to determine the distance at which intensity is maximized, leading to a ``bi-level'' optimization problem for $\rho$~\cite{blondel2022efficient}; while such optimization problems are potentially tractable, it is more efficient if we can avoid nested optimization entirely. To estimate the focal position at frequency~$f$, we instead compute the weighted average:
\begin{equation}
    z(\rho, f) = \arg \max_z |\textbf{E}(z)|^2 \approx \frac{\int_a^b z|\textbf{E}(z)|^2dz}{\int_a^b |\textbf{E}(z))|^2dz} \label{eq:argmax_approx}
\end{equation}
for some interval $(a,b)$ around $z_0$. Each integral can be approximated using (e.g.) a Riemann sum; hence, the position $z$ is approximated as the average of a set of positions over the interval, weighted by a normalized intensity at each position.

As discussed above, we also wish to ensure that we do not sacrifice too much focal intensity to maximize shifting, so we impose a constraint
\begin{equation}
    |\textbf{E}(z_0)|^2 \ge \alpha I_0 \quad \text{at frequencies} \ f_0 \pm\delta f
\end{equation}
for some hyperparameter $\alpha<1$; we used $\alpha = 0.9$.  (Alternatively, one could constrain $|\textbf{E}(z(\rho, f_0 \pm\delta f)|^2$, but if $\delta f$ is sufficiently small this should be nearly equivalent.)  This constraint also helps to ensure that the approximation in Eq.~\ref{eq:argmax_approx} is valid, by requiring high-intensity focal peak.

This approach enables us to estimate the maximal attainable focal shift. We perform this optimization twice, for both  signs in Eq.~\ref{Eq:max_shift}, to determine the maximal shift in both directions. Combining the results from these two optimizations provides a reasonable magnitude for the possible focal shift in both directions, as shown in the Results section below.

\subsubsection{Stage three: Designing matched-shift metalenses}
After we determined a reasonable shift magnitude, we can simply maximize the field intensity at $\pm$shifts of up to that magnitude for the corresponding frequencies.   To maximize all shift intensities simultaneously, we formulate the objective as a ``max-min" problem, i.e. $\max_\rho \{\min_k |\mathbf{E}(z_k,f_k)|^2\}$, which can then be transformed into a differentiable form using an epigraph reformulation~\cite{boyd2004convex}:
\begin{equation}
    \begin{split}
    \max_\rho \quad & t \\    \mathrm{s.t.} \quad   t\ \le \ & |\mathbf{E}_k(z_k, f_k)|^2
\end{split}
\end{equation}
where we introduced an additional auxiliary variable $t \in \mathbb{R}$, and $\mathbf{E}(z_k, f_k)$ is the electric field of frequencies $f_k$ at points $z_k$ discussed below.

\begin{figure}[ht!]
    \centering
    \includegraphics[width=0.8\textwidth]{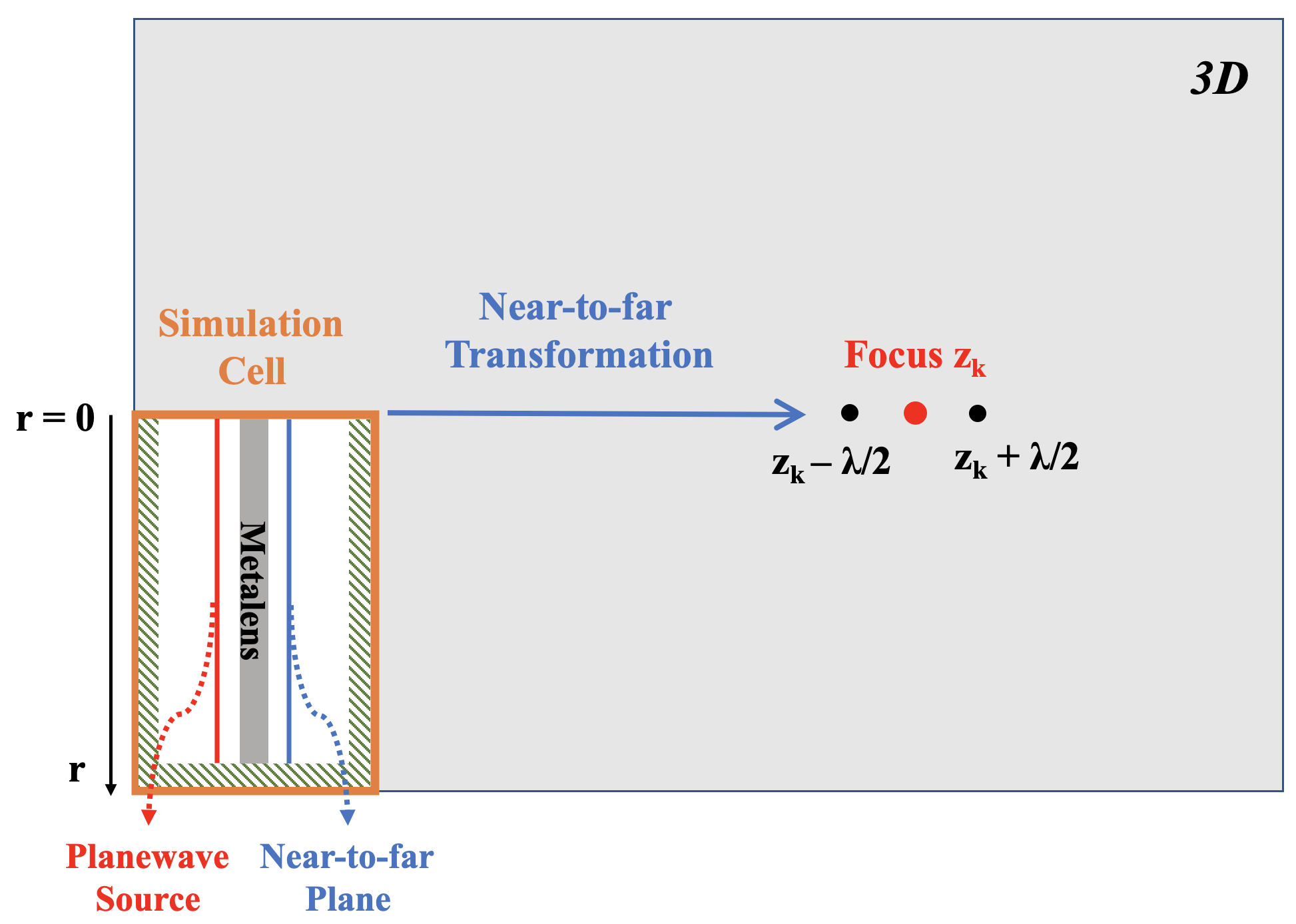}
    \caption{Schematic of the computational domain and optimization problem. The simulation cell (orange box) is a cross section of a 3D axisymmetric problem, whose transmitted fields are propagated by a near-to-far--field transformation to a focal point $z_k$.  Different $z_k$ are targeted depending on wavelength~$\lambda$, and adjacent spots $z_k \pm \lambda/2$ are monitored to ensure a focal peak at $z_k$.  The metalens region (gray) is designed by freeform topology optimization subject to fabrication constraints.}
    \label{fig:2Dplot}
\end{figure}

However, simply maximizing intensity at a point does not ensure that the point will be a \emph{maximum} intensity (a focus), especially if many foci are being maximized simultaneously.  To ensure that each point $z_k$ is a local maximum (at $f_k$), we impose additional constraints:
\begin{equation}
\begin{split}
    |\mathbf{E}(z_k\pm \frac{\lambda_k}{2}, f_k)|^2 \le |\mathbf{E}_(z_k, f_k)|^2 
\end{split}
\end{equation}
where $\lambda_k = \frac{c}{f_k}$ is the wavelength at frequency $f_k$. The constraint ensures that the focus will be within half a wavelength of~$z_k$. A schematic is shown in Fig.~\ref{fig:2Dplot}.  

Finally, we ideally should prescribe foci at a sufficient number of frequencies within the operating range to ensure that the focal spot shifts linearly with frequency. However, we find empirically that simply optimizing the foci at the \emph{endpoints} of the frequency range typically suffices to ensure that intermediate frequencies shift the focal spot nearly linearly.  Therefore, to reduce the computational cost, in this paper we simply optimize at $z_k = z_0 \pm \text{shift}$.  In future work, more precise linearity could be imposed by  including more foci at intermediate frequencies.  

\section{Results}

Using our formulation, we designed separate refractive and diffractive metalenses that shift focus as the frequency increases from 250~GHz to 300~GHz. Both metalenses are made of a typical 3D printing material with $\varepsilon = 2.66$ and a loss tangent of 0.03, with negligible material dispersion over this bandwidth~\cite{Chan20}. The design region has a radius of 12~mm (roughly 10 wavelengths) and a height of 2~mm, and it rests on a solid substrate of the same material with a thickness of 0.5~mm. We imposed a minimum lengthscale of 0.15~mm, consistent with the parameters from Ref.~\citenum{Chan20}. 

\begin{figure}[ht!]
    \centering
    \includegraphics[width=\textwidth]{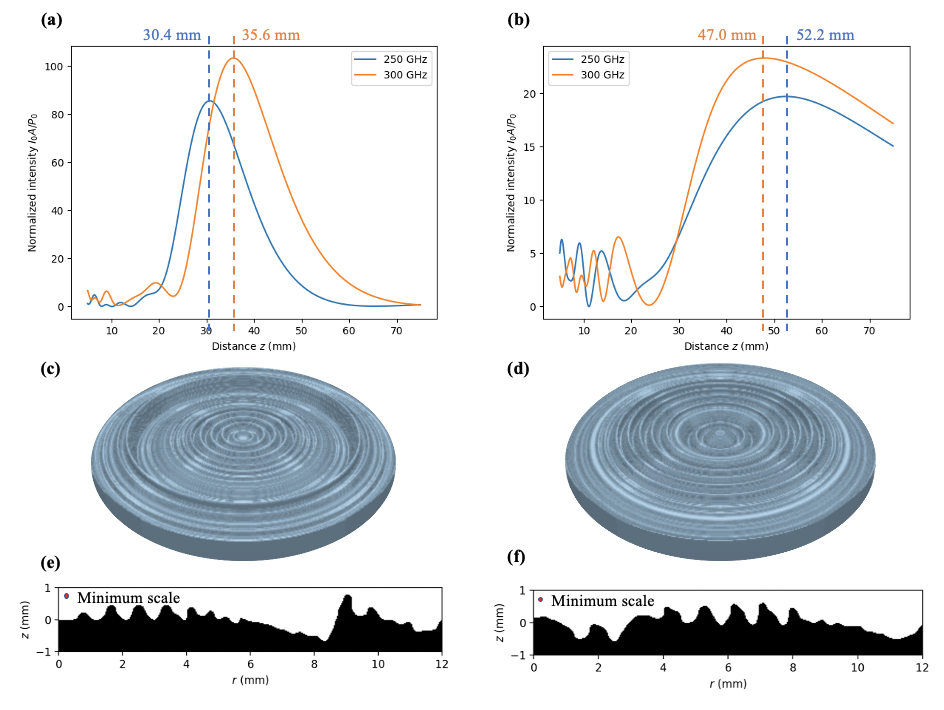}
    \caption{Performance of ``diffractive'' (focal length increases with frequency) and ``refractive'' (decreases) metalenses: Normalized field intensity along the vertical $z$-axis of (a) diffractive metalens and (b) refractive metalens; 3D structure of (c) the diffractive metalens and (d) the refractive metalens along with corresponding 2D cross sections (e) and (f), respectively, where the imposed minimum fabrication lengthscale (0.15~mm) is indicated by the diameter of a red dot.}
    \label{fig:results}
\end{figure}

Following the scanning-optimization procedure described in the previous section, we found that for a frequency shift over 250--300~GHz, the maximal focal shift is approximately 5~mm. Because of the analogy between metalens and diffractive Fresnel lenses\cite{Engelberg2020,Khorasaninejad17}, we expected that ``diffractive'' metalenses (focal length increasing with frequency) should generally be easier to achieve compared to ``refractive'' metalenses (focal length decreasing with frequency), so we chose a slightly longer focal length (smaller numerical aperture) for the latter case.  We thus targeted focal locations of 30--35~mm for the diffractive metalens and 50--45~mm for the refractive metalens. Furthermore, we conducted a parameter sweep of the target focal length, examining a range of target focal spots near those values to identify those that exhibited the highest-quality focus. For every optimization, we initialized the structure to be solid in the bottom half of the design region and air in the top half. We started with $\beta = 8$ in the projection Eq.~\ref{eq:tanh_proj} and doubled it after every epoch of 200 optimization iterations, for a total of four epochs. Initially, no fabrication constraints were included other than the penalty on non-binary pixels. Connectivity constraints were introduced after the second epoch, and lengthscale constraints were added in the fourth epoch.

\begin{figure}[htb!]
    \centering
    \includegraphics[width=0.6\textwidth]{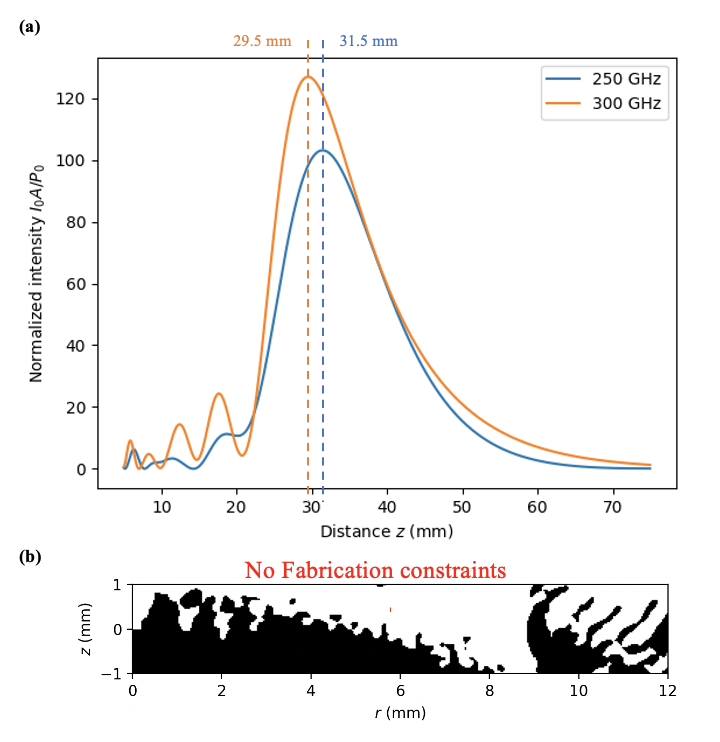}
    \caption{Refractive metalens without fabrication constraints (neither a minimum lengthscale nor connectivity were imposed): (a) normalized field intensity along the vertical $z$-axis; (b) cross section of the structure.}
    \label{fig:noconstraint}
\end{figure}

We achieved a diffractive metalens that focused at $z = 30.4$~mm at 250~GHz and at $z=35.6$~mm at 300~GHz, and a refractive metalens that focused at $z = 47.0$~mm at 250~GHz and at $z=52.2$~mm at 300~GHz. We computed the normalized field intensity $\frac{I_0A}{P_0}$, where $A$ is the area of the lens and $P_0$ is the transmitted power measured right above the metalens. The performance and structures of final designs are shown in Fig.~\ref{fig:results}. We observed that designing the refractive metalens indeed proved more challenging than the diffractive metalens. The refractive metalens had a lower focal quality compared to the diffractive metalens: the numerical aperture was around 0.22 compared to 0.35, and the Strehl ratio was around 0.38 compared to 0.57. However, we are not aware of any \emph{fundamental} theoretical limitations affecting the performance of refractive metalenses, and this difference seems largely due to the imposed manufacturing constraints. In fact, we found that with more lenient fabrication constraints and less ambitious focus shift ranges, we could improve the performance of the ``refractive'' metalens. As shown in Fig.~\ref{fig:noconstraint}, we designed a refractive metalens without fabrication constraints that shifts focus from $z=31.5$~mm to $z=29.5$ when frequency increased from 250~GHz to 300~GHz. Its numerical aperture was around 0.36 with a Strehl ratio of 0.54.

\section{Conclusion}
Previous research using traditional methods for designing focusing metalenses generally resulted in ``diffractive'' structures, for which the focal length increases with frequency. In this study, we employed topology optimization that allows us to pair a diffractive metalens with a complementary ``refractive'' metalens (for which the focal length decreases with frequency).   Not only could such a pair be practically useful for THz imaging, but it also illustrates an interesting physical dichotomy between the difficulty of achieving the two distinct dispersion behaviors in a thin structure, even with optimized freeform designs.   A key factor in making such design tractable is careful formulation of the optimization problem, in order to make precise the notion of a ``maximally shifted focus'' without imposing impractical computational costs, and in our case this turned out to involve three coupled stages of optimization problems.

An open theoretical challenge would be to rigorously quantify the apparent asymmetry of the design problem: In what sense is the refractive direction ``harder'' than the diffractive direction?  Perhaps one could build off recent progress in fundamental bounds on scattering processes~\cite{Chao2022}; however, we empirically find that refractive designs appears to be more limited by fabrication constraints, which have thus far been challenging to incorporate into theoretical bounds~\cite{Chao2022}.  More generally, we believe that freeform volumetric design of 3D-printed metalenses for microwave and terahertz applications is ripe for future experimental and theoretical investigation, because it pairs a vast number of experimental degrees of freedom with tractable computational domains ($< 100\lambda$ in diameter) for fullwave simulation and optimization.

\section*{Funding}

MC and SGJ were supported in part by a grant from the Simons Foundation.  AH was supported by the Department of Defense (DoD) through the National Defense Science \& Engineering Graduate (NDSEG) Fellowship Program.  KFC and CHC were supported in part by the Hong Kong Research Grants Council General Research Fund under the project CityU 11217720..

\section*{Notes}

The authors declare no competing financial interest.

\section*{Acknowledgments}

We are grateful to Aristeidis Karalis at MIT for helpful discussions.

\bibliography{ref}

\providecommand{\latin}[1]{#1}
\makeatletter
\providecommand{\doi}
  {\begingroup\let\do\@makeother\dospecials
  \catcode`\{=1 \catcode`\}=2 \doi@aux}
\providecommand{\doi@aux}[1]{\endgroup\texttt{#1}}
\makeatother
\providecommand*\mcitethebibliography{\thebibliography}
\csname @ifundefined\endcsname{endmcitethebibliography}  {\let\endmcitethebibliography\endthebibliography}{}
\begin{mcitethebibliography}{50}
\providecommand*\natexlab[1]{#1}
\providecommand*\mciteSetBstSublistMode[1]{}
\providecommand*\mciteSetBstMaxWidthForm[2]{}
\providecommand*\mciteBstWouldAddEndPuncttrue
  {\def\EndOfBibitem{\unskip.}}
\providecommand*\mciteBstWouldAddEndPunctfalse
  {\let\EndOfBibitem\relax}
\providecommand*\mciteSetBstMidEndSepPunct[3]{}
\providecommand*\mciteSetBstSublistLabelBeginEnd[3]{}
\providecommand*\EndOfBibitem{}
\mciteSetBstSublistMode{f}
\mciteSetBstMaxWidthForm{subitem}{(\alph{mcitesubitemcount})}
\mciteSetBstSublistLabelBeginEnd
  {\mcitemaxwidthsubitemform\space}
  {\relax}
  {\relax}

\bibitem[Stone and George(1988)Stone, and George]{Stone88}
Stone,~T.; George,~N. Hybrid diffractive-refractive lenses and achromats. \emph{Appl. Opt.} \textbf{1988}, \emph{27}, 2960--2971\relax
\mciteBstWouldAddEndPuncttrue
\mciteSetBstMidEndSepPunct{\mcitedefaultmidpunct}
{\mcitedefaultendpunct}{\mcitedefaultseppunct}\relax
\EndOfBibitem
\bibitem[Engelberg and Levy(2020)Engelberg, and Levy]{Engelberg2020}
Engelberg,~J.; Levy,~U. The advantages of metalenses over diffractive lenses. \emph{Nature communications} \textbf{2020}, \emph{11}, 1991--4\relax
\mciteBstWouldAddEndPuncttrue
\mciteSetBstMidEndSepPunct{\mcitedefaultmidpunct}
{\mcitedefaultendpunct}{\mcitedefaultseppunct}\relax
\EndOfBibitem
\bibitem[Khorasaninejad and Capasso(20171201)Khorasaninejad, and Capasso]{Khorasaninejad17}
Khorasaninejad,~M.; Capasso,~F. Metalenses: Versatile multifunctional photonic components. \emph{Science.} \textbf{20171201}, \emph{358}, 1146--1146\relax
\mciteBstWouldAddEndPuncttrue
\mciteSetBstMidEndSepPunct{\mcitedefaultmidpunct}
{\mcitedefaultendpunct}{\mcitedefaultseppunct}\relax
\EndOfBibitem
\bibitem[Born and Wolf(2013)Born, and Wolf]{born2013principles}
Born,~M.; Wolf,~E. \emph{Principles of optics: electromagnetic theory of propagation, interference and diffraction of light}; Elsevier, 2013\relax
\mciteBstWouldAddEndPuncttrue
\mciteSetBstMidEndSepPunct{\mcitedefaultmidpunct}
{\mcitedefaultendpunct}{\mcitedefaultseppunct}\relax
\EndOfBibitem
\bibitem[Miyamoto and Miyamoto(1961-01-01)Miyamoto, and Miyamoto]{PFL1961}
Miyamoto,~K.; Miyamoto,~K. The Phase Fresnel Lens. \emph{Journal of the Optical Society of America.} \textbf{1961-01-01}, \emph{51}, 17--\relax
\mciteBstWouldAddEndPuncttrue
\mciteSetBstMidEndSepPunct{\mcitedefaultmidpunct}
{\mcitedefaultendpunct}{\mcitedefaultseppunct}\relax
\EndOfBibitem
\bibitem[Chao \latin{et~al.}(2022)Chao, Strekha, Kuate~Defo, Molesky, and Rodriguez]{Chao2022}
Chao,~P.; Strekha,~B.; Kuate~Defo,~R.; Molesky,~S.; Rodriguez,~A.~W. Physical limits in electromagnetism. \emph{Nature Reviews Physics} \textbf{2022}, \emph{4}, 543–559\relax
\mciteBstWouldAddEndPuncttrue
\mciteSetBstMidEndSepPunct{\mcitedefaultmidpunct}
{\mcitedefaultendpunct}{\mcitedefaultseppunct}\relax
\EndOfBibitem
\bibitem[Siegel(2002)]{Siegel02}
Siegel,~P. Terahertz technology. \emph{IEEE Transactions on Microwave Theory and Techniques} \textbf{2002}, \emph{50}, 910--928\relax
\mciteBstWouldAddEndPuncttrue
\mciteSetBstMidEndSepPunct{\mcitedefaultmidpunct}
{\mcitedefaultendpunct}{\mcitedefaultseppunct}\relax
\EndOfBibitem
\bibitem[Song and Nagatsuma(2011)Song, and Nagatsuma]{Song11}
Song,~H.-J.; Nagatsuma,~T. Present and Future of Terahertz Communications. \emph{IEEE Transactions on Terahertz Science and Technology} \textbf{2011}, \emph{1}, 256--263\relax
\mciteBstWouldAddEndPuncttrue
\mciteSetBstMidEndSepPunct{\mcitedefaultmidpunct}
{\mcitedefaultendpunct}{\mcitedefaultseppunct}\relax
\EndOfBibitem
\bibitem[Wu \latin{et~al.}(2020)Wu, Chan, Qu, and Chan]{Chan20}
Wu,~G.-B.; Chan,~K.~F.; Qu,~S.-W.; Chan,~C.~H. A 2-D Beam-Scanning Bessel Launcher for Terahertz Applications. \emph{IEEE Transactions on Antennas and Propagation} \textbf{2020}, \emph{68}, 5893--5903\relax
\mciteBstWouldAddEndPuncttrue
\mciteSetBstMidEndSepPunct{\mcitedefaultmidpunct}
{\mcitedefaultendpunct}{\mcitedefaultseppunct}\relax
\EndOfBibitem
\bibitem[Wu \latin{et~al.}(2022)Wu, Zeng, Chan, Qu, Shaw, and Chan]{Chan22}
Wu,~G.-B.; Zeng,~Y.-S.; Chan,~K.~F.; Qu,~S.-W.; Shaw,~J.; Chan,~C.~H. 3-D Printed 3-D Near-Field Focus-Scanning Lens for Terahertz Applications. \emph{IEEE Transactions on Antennas and Propagation} \textbf{2022}, \emph{70}, 10007--10016\relax
\mciteBstWouldAddEndPuncttrue
\mciteSetBstMidEndSepPunct{\mcitedefaultmidpunct}
{\mcitedefaultendpunct}{\mcitedefaultseppunct}\relax
\EndOfBibitem
\bibitem[Liu \latin{et~al.}(2023)Liu, Yang, Xu, and Li]{liu23}
Liu,~C.; Yang,~F.; Xu,~S.; Li,~M. Reconfigurable Metasurface: A Systematic Categorization and Recent Advances. \emph{Electromagnetic Science} \textbf{2023}, \emph{1}, 1--23\relax
\mciteBstWouldAddEndPuncttrue
\mciteSetBstMidEndSepPunct{\mcitedefaultmidpunct}
{\mcitedefaultendpunct}{\mcitedefaultseppunct}\relax
\EndOfBibitem
\bibitem[Yi \latin{et~al.}(2017)Yi, Qu, Chen, Bai, Ng, and Chan]{Yi2017FlatTR}
Yi,~H.; Qu,~S.; Chen,~B.~J.; Bai,~X.; Ng,~K.~B.; Chan,~C.~H. Flat Terahertz Reflective Focusing Metasurface with Scanning Ability. \emph{Scientific Reports} \textbf{2017}, \emph{7}\relax
\mciteBstWouldAddEndPuncttrue
\mciteSetBstMidEndSepPunct{\mcitedefaultmidpunct}
{\mcitedefaultendpunct}{\mcitedefaultseppunct}\relax
\EndOfBibitem
\bibitem[Yu and Capasso(2014)Yu, and Capasso]{yu2014flat}
Yu,~N.; Capasso,~F. Flat optics with designer metasurfaces. \emph{Nature materials} \textbf{2014}, \emph{13}, 139--150\relax
\mciteBstWouldAddEndPuncttrue
\mciteSetBstMidEndSepPunct{\mcitedefaultmidpunct}
{\mcitedefaultendpunct}{\mcitedefaultseppunct}\relax
\EndOfBibitem
\bibitem[Beneck \latin{et~al.}(2024)Beneck, Kang, Jenkins, Campbell, and Werner]{Beneck24}
Beneck,~R.~J.; Kang,~L.; Jenkins,~R.~P.; Campbell,~S.~D.; Werner,~D.~H. Superscattering of electromagnetic waves from subwavelength dielectric structures. \emph{Opt. Express} \textbf{2024}, \emph{32}, 19410--19423\relax
\mciteBstWouldAddEndPuncttrue
\mciteSetBstMidEndSepPunct{\mcitedefaultmidpunct}
{\mcitedefaultendpunct}{\mcitedefaultseppunct}\relax
\EndOfBibitem
\bibitem[Chen \latin{et~al.}(2018)Chen, Zhu, Sanjeev, Khorasaninejad, Shi, Lee, and Capasso]{chen2018broadband}
Chen,~W.~T.; Zhu,~A.~Y.; Sanjeev,~V.; Khorasaninejad,~M.; Shi,~Z.; Lee,~E.; Capasso,~F. A broadband achromatic metalens for focusing and imaging in the visible. \emph{Nature nanotechnology} \textbf{2018}, \emph{13}, 220--226\relax
\mciteBstWouldAddEndPuncttrue
\mciteSetBstMidEndSepPunct{\mcitedefaultmidpunct}
{\mcitedefaultendpunct}{\mcitedefaultseppunct}\relax
\EndOfBibitem
\bibitem[Cui \latin{et~al.}(2024)Cui, Zhang, Alù, Wegener, Pendry, Luo, Lai, Wang, Lin, Chen, Chen, Wu, Yin, Zhao, Chen, Li, Zhou, Engheta, Asadchy, Simovski, Tretyakov, Yang, Campbell, Hao, Werner, Sun, Zhou, Xu, Sun, Zhou, Li, Zheng, Chen, Li, Zhu, Zhou, Zhao, Liu, Zhang, Zhang, Gu, Xiao, Liu, Zhang, Tang, Li, Zentgraf, Koshelev, Kivshar, Li, Badloe, Huang, Rho, Wang, Tsai, Bykov, Krasavin, Zayats, McDonnell, Ellenbogen, Luo, Pu, Garcia-Vidal, Liu, Li, Tang, Ma, Zhang, Luo, Zhang, Zhang, He, Zhang, Wan, Wu, Liu, Jiang, Zhang, Qiu, Ma, Liu, Li, Han, Li, Cotrufo, Caloz, Deck-Léger, Bahrami, Céspedes, Galiffi, Huidobro, Cheng, Dai, Ke, Zhang, Galdi, and di~Renzo]{Cui_2024}
Cui,~T.~J. \latin{et~al.}  Roadmap on electromagnetic metamaterials and metasurfaces. \emph{Journal of Physics: Photonics} \textbf{2024}, \emph{6}, 032502\relax
\mciteBstWouldAddEndPuncttrue
\mciteSetBstMidEndSepPunct{\mcitedefaultmidpunct}
{\mcitedefaultendpunct}{\mcitedefaultseppunct}\relax
\EndOfBibitem
\bibitem[Kuznetsov \latin{et~al.}(2024)Kuznetsov, Brongersma, Yao, Chen, Levy, Tsai, Zheludev, Faraon, Arbabi, Yu, Chanda, Crozier, Kildishev, Wang, Yang, Valentine, Genevet, Fan, Miller, Majumdar, Fr{\"o}ch, Brady, Heide, Veeraraghavan, Engheta, Alù, Polman, Atwater, Thureja, Paniagua-Dominguez, Ha, Barreda, Schuller, Staude, Grinblat, Kivshar, Peana, Yelin, Senichev, Shalaev, Saha, Boltasseva, Rho, Oh, Kim, Park, Devlin, and Pala]{Kuznetsov24}
Kuznetsov,~A.~I. \latin{et~al.}  Roadmap for Optical Metasurfaces. \emph{ACS Photonics} \textbf{2024}, \emph{11}, 816--865, PMID: 38550347\relax
\mciteBstWouldAddEndPuncttrue
\mciteSetBstMidEndSepPunct{\mcitedefaultmidpunct}
{\mcitedefaultendpunct}{\mcitedefaultseppunct}\relax
\EndOfBibitem
\bibitem[Wang \latin{et~al.}(2023)Wang, Yu, Phan, Dhuey, and Fan]{wang23}
Wang,~E.~W.; Yu,~S.-J.; Phan,~T.; Dhuey,~S.; Fan,~J.~A. Arbitrary Achromatic Polarization Control with Reconfigurable Metasurface Systems. \emph{Laser \& Photonics Reviews} \textbf{2023}, \emph{17}, 2200926\relax
\mciteBstWouldAddEndPuncttrue
\mciteSetBstMidEndSepPunct{\mcitedefaultmidpunct}
{\mcitedefaultendpunct}{\mcitedefaultseppunct}\relax
\EndOfBibitem
\bibitem[Roberts \latin{et~al.}(2023)Roberts, Ballew, Zheng, Garcia, Camayd-Mu{\~n}oz, Hon, and Faraon]{roberts20233d}
Roberts,~G.; Ballew,~C.; Zheng,~T.; Garcia,~J.~C.; Camayd-Mu{\~n}oz,~S.; Hon,~P.~W.; Faraon,~A. 3D-patterned inverse-designed mid-infrared metaoptics. \emph{Nature Communications} \textbf{2023}, \emph{14}, 2768\relax
\mciteBstWouldAddEndPuncttrue
\mciteSetBstMidEndSepPunct{\mcitedefaultmidpunct}
{\mcitedefaultendpunct}{\mcitedefaultseppunct}\relax
\EndOfBibitem
\bibitem[Pestourie \latin{et~al.}(2018)Pestourie, P\'{e}rez-Arancibia, Lin, Shin, Capasso, and Johnson]{Pestourie18}
Pestourie,~R.; P\'{e}rez-Arancibia,~C.; Lin,~Z.; Shin,~W.; Capasso,~F.; Johnson,~S.~G. Inverse design of large-area metasurfaces. \emph{Opt. Express} \textbf{2018}, \emph{26}, 33732--33747\relax
\mciteBstWouldAddEndPuncttrue
\mciteSetBstMidEndSepPunct{\mcitedefaultmidpunct}
{\mcitedefaultendpunct}{\mcitedefaultseppunct}\relax
\EndOfBibitem
\bibitem[Lin \latin{et~al.}(2019)Lin, Liu, Pestourie, and Johnson]{Lin19}
Lin,~Z.; Liu,~V.; Pestourie,~R.; Johnson,~S.~G. Topology optimization of freeform large-area metasurfaces. \emph{Opt. Express} \textbf{2019}, \emph{27}, 15765--15775\relax
\mciteBstWouldAddEndPuncttrue
\mciteSetBstMidEndSepPunct{\mcitedefaultmidpunct}
{\mcitedefaultendpunct}{\mcitedefaultseppunct}\relax
\EndOfBibitem
\bibitem[Popov \latin{et~al.}(2019)Popov, Yakovleva, Boust, Pelouard, Pardo, and Burokur]{LPA2019}
Popov,~V.; Yakovleva,~M.; Boust,~F.; Pelouard,~J.-L.; Pardo,~F.; Burokur,~S.~N. Designing Metagratings via Local Periodic Approximation: From Microwaves to Infrared. \emph{Phys. Rev. Appl.} \textbf{2019}, \emph{11}, 044054\relax
\mciteBstWouldAddEndPuncttrue
\mciteSetBstMidEndSepPunct{\mcitedefaultmidpunct}
{\mcitedefaultendpunct}{\mcitedefaultseppunct}\relax
\EndOfBibitem
\bibitem[Christiansen \latin{et~al.}(2020)Christiansen, Lin, Roques-Carmes, Salamin, Kooi, Joannopoulos, Solja\v{c}i\'{c}, and Johnson]{Christiansen20}
Christiansen,~R.~E.; Lin,~Z.; Roques-Carmes,~C.; Salamin,~Y.; Kooi,~S.~E.; Joannopoulos,~J.~D.; Solja\v{c}i\'{c},~M.; Johnson,~S.~G. Fullwave {Maxwell} inverse design of axisymmetric, tunable, and multi-scale multi-wavelength metalenses. \emph{Opt. Express} \textbf{2020}, \emph{28}, 33854--33868\relax
\mciteBstWouldAddEndPuncttrue
\mciteSetBstMidEndSepPunct{\mcitedefaultmidpunct}
{\mcitedefaultendpunct}{\mcitedefaultseppunct}\relax
\EndOfBibitem
\bibitem[Lin \latin{et~al.}(2021)Lin, Roques-Carmes, Christiansen, Solja{\v{c}}i{\'c}, and Johnson]{lin2021computational}
Lin,~Z.; Roques-Carmes,~C.; Christiansen,~R.~E.; Solja{\v{c}}i{\'c},~M.; Johnson,~S.~G. Computational inverse design for ultra-compact single-piece metalenses free of chromatic and angular aberration. \emph{Applied Physics Letters} \textbf{2021}, \emph{118}\relax
\mciteBstWouldAddEndPuncttrue
\mciteSetBstMidEndSepPunct{\mcitedefaultmidpunct}
{\mcitedefaultendpunct}{\mcitedefaultseppunct}\relax
\EndOfBibitem
\bibitem[Sigmund and Maute(2013)Sigmund, and Maute]{Maute13}
Sigmund,~O.; Maute,~K. Topology optimization approaches: A comparative review. \emph{Structural and Multidisciplinary Optimization} \textbf{2013}, \emph{48}, 1031--1055\relax
\mciteBstWouldAddEndPuncttrue
\mciteSetBstMidEndSepPunct{\mcitedefaultmidpunct}
{\mcitedefaultendpunct}{\mcitedefaultseppunct}\relax
\EndOfBibitem
\bibitem[Hammond \latin{et~al.}(2021)Hammond, Oskooi, Johnson, and Ralph]{Hammond21DRC}
Hammond,~A.; Oskooi,~A.; Johnson,~S.; Ralph,~S. Photonic topology optimization with semiconductor-foundry design-rule constraints. \emph{Optics Express} \textbf{2021}, \emph{29}\relax
\mciteBstWouldAddEndPuncttrue
\mciteSetBstMidEndSepPunct{\mcitedefaultmidpunct}
{\mcitedefaultendpunct}{\mcitedefaultseppunct}\relax
\EndOfBibitem
\bibitem[Christiansen(2023)]{Christiansen_2023}
Christiansen,~R.~E. Inverse design of optical mode converters by topology optimization: tutorial. \emph{Journal of Optics} \textbf{2023}, \emph{25}, 083501\relax
\mciteBstWouldAddEndPuncttrue
\mciteSetBstMidEndSepPunct{\mcitedefaultmidpunct}
{\mcitedefaultendpunct}{\mcitedefaultseppunct}\relax
\EndOfBibitem
\bibitem[Jensen and Sigmund(2011)Jensen, and Sigmund]{jensen2011topology}
Jensen,~J.~S.; Sigmund,~O. Topology optimization for nano-photonics. \emph{Laser \& Photonics Reviews} \textbf{2011}, \emph{5}, 308--321\relax
\mciteBstWouldAddEndPuncttrue
\mciteSetBstMidEndSepPunct{\mcitedefaultmidpunct}
{\mcitedefaultendpunct}{\mcitedefaultseppunct}\relax
\EndOfBibitem
\bibitem[Berenger(1994)]{BEREGNER_1994}
Berenger,~J.-P. A Perfectly Matched Layer for the Absorption of Electromagnetic Waves. \emph{Journal of Computational Physics} \textbf{1994}, \emph{114}, 185--200\relax
\mciteBstWouldAddEndPuncttrue
\mciteSetBstMidEndSepPunct{\mcitedefaultmidpunct}
{\mcitedefaultendpunct}{\mcitedefaultseppunct}\relax
\EndOfBibitem
\bibitem[Sacks \latin{et~al.}(1995)Sacks, Kingsland, Lee, and Lee]{Sacks1995PML}
Sacks,~Z.; Kingsland,~D.; Lee,~R.; Lee,~J.-F. A perfectly matched anisotropic absorber for use as an absorbing boundary condition. \emph{IEEE Transactions on Antennas and Propagation} \textbf{1995}, \emph{43}, 1460--1463\relax
\mciteBstWouldAddEndPuncttrue
\mciteSetBstMidEndSepPunct{\mcitedefaultmidpunct}
{\mcitedefaultendpunct}{\mcitedefaultseppunct}\relax
\EndOfBibitem
\bibitem[Oskooi and Johnson(2011)Oskooi, and Johnson]{OSKOOI2011PML}
Oskooi,~A.; Johnson,~S.~G. Distinguishing correct from incorrect {PML} proposals and a corrected unsplit {PML} for anisotropic, dispersive media. \emph{Journal of Computational Physics} \textbf{2011}, \emph{230}, 2369--2377\relax
\mciteBstWouldAddEndPuncttrue
\mciteSetBstMidEndSepPunct{\mcitedefaultmidpunct}
{\mcitedefaultendpunct}{\mcitedefaultseppunct}\relax
\EndOfBibitem
\bibitem[Hammond \latin{et~al.}(2022)Hammond, Oskooi, Chen, Lin, Johnson, and Ralph]{Hammond2022}
Hammond,~A.~M.; Oskooi,~A.; Chen,~M.; Lin,~Z.; Johnson,~S.~G.; Ralph,~S.~E. High-performance hybrid time/frequency-domain topology optimization for large-scale photonics inverse design. \emph{Optics Express} \textbf{2022}, \emph{30}, 4467\relax
\mciteBstWouldAddEndPuncttrue
\mciteSetBstMidEndSepPunct{\mcitedefaultmidpunct}
{\mcitedefaultendpunct}{\mcitedefaultseppunct}\relax
\EndOfBibitem
\bibitem[Oskooi \latin{et~al.}(2010)Oskooi, Roundy, Ibanescu, Bermel, Joannopoulos, and Johnson]{oskooi2010meep}
Oskooi,~A.~F.; Roundy,~D.; Ibanescu,~M.; Bermel,~P.; Joannopoulos,~J.~D.; Johnson,~S.~G. MEEP: A flexible free-software package for electromagnetic simulations by the {FDTD} method. \emph{Computer Physics Communications} \textbf{2010}, \emph{181}, 687--702\relax
\mciteBstWouldAddEndPuncttrue
\mciteSetBstMidEndSepPunct{\mcitedefaultmidpunct}
{\mcitedefaultendpunct}{\mcitedefaultseppunct}\relax
\EndOfBibitem
\bibitem[Sigmund and Petersson(1998)Sigmund, and Petersson]{sigmund1998numerical}
Sigmund,~O.; Petersson,~J. Numerical instabilities in topology optimization: a survey on procedures dealing with checkerboards, mesh-dependencies and local minima. \emph{Structural optimization} \textbf{1998}, \emph{16}, 68--75\relax
\mciteBstWouldAddEndPuncttrue
\mciteSetBstMidEndSepPunct{\mcitedefaultmidpunct}
{\mcitedefaultendpunct}{\mcitedefaultseppunct}\relax
\EndOfBibitem
\bibitem[Sigmund(2007)]{sigmund2007morphology}
Sigmund,~O. Morphology-based black and white filters for topology optimization. \emph{Structural and Multidisciplinary Optimization} \textbf{2007}, \emph{33}, 401--424\relax
\mciteBstWouldAddEndPuncttrue
\mciteSetBstMidEndSepPunct{\mcitedefaultmidpunct}
{\mcitedefaultendpunct}{\mcitedefaultseppunct}\relax
\EndOfBibitem
\bibitem[Wang \latin{et~al.}(2011)Wang, Lazarov, and Sigmund]{Wang2011}
Wang,~F.; Lazarov,~B.~S.; Sigmund,~O. On projection methods, convergence and robust formulations in topology optimization. \emph{Structural and Multidisciplinary Optimization} \textbf{2011}, \emph{43}, 767--784\relax
\mciteBstWouldAddEndPuncttrue
\mciteSetBstMidEndSepPunct{\mcitedefaultmidpunct}
{\mcitedefaultendpunct}{\mcitedefaultseppunct}\relax
\EndOfBibitem
\bibitem[Lazarov \latin{et~al.}(2016)Lazarov, Wang, and Sigmund]{lazarov2016length}
Lazarov,~B.~S.; Wang,~F.; Sigmund,~O. Length scale and manufacturability in density-based topology optimization. \emph{Archive of Applied Mechanics} \textbf{2016}, \emph{86}, 189--218\relax
\mciteBstWouldAddEndPuncttrue
\mciteSetBstMidEndSepPunct{\mcitedefaultmidpunct}
{\mcitedefaultendpunct}{\mcitedefaultseppunct}\relax
\EndOfBibitem
\bibitem[Wu \latin{et~al.}(2019)Wu, Zeng, Chan, Qu, and Chan]{Wu2019}
Wu,~G.-B.; Zeng,~Y.-S.; Chan,~K.~F.; Qu,~S.-W.; Chan,~C.~H. 3-D Printed Circularly Polarized Modified Fresnel Lens Operating at Terahertz Frequencies. \emph{IEEE Transactions on Antennas and Propagation} \textbf{2019}, \emph{67}, 4429--4437\relax
\mciteBstWouldAddEndPuncttrue
\mciteSetBstMidEndSepPunct{\mcitedefaultmidpunct}
{\mcitedefaultendpunct}{\mcitedefaultseppunct}\relax
\EndOfBibitem
\bibitem[Taflove and Hagness(2000)Taflove, and Hagness]{Taflove2005}
Taflove,~A.; Hagness,~S.~C. \emph{Computational electrodynamics : the finite-difference time-domain method.}, 2nd ed.; Artech House antennas and propagation library; Artech House: Boston, 2000\relax
\mciteBstWouldAddEndPuncttrue
\mciteSetBstMidEndSepPunct{\mcitedefaultmidpunct}
{\mcitedefaultendpunct}{\mcitedefaultseppunct}\relax
\EndOfBibitem
\bibitem[Oskooi and Johnson(2013)Oskooi, and Johnson]{oskooi2013}
Oskooi,~A.; Johnson,~S.~G. In \emph{Advances in {FDTD} Computational Electrodynamics: Photonics and Nanotechnology}; Taflove,~A., Oskooi,~A., Johnson,~S.~G., Eds.; Artech: Boston, 2013; Chapter 4, pp 65--100\relax
\mciteBstWouldAddEndPuncttrue
\mciteSetBstMidEndSepPunct{\mcitedefaultmidpunct}
{\mcitedefaultendpunct}{\mcitedefaultseppunct}\relax
\EndOfBibitem
\bibitem[Niederberger \latin{et~al.}(2014)Niederberger, Fattal, Gauger, Fan, and Beausoleil]{niederberger2014sensitivity}
Niederberger,~A.~C.; Fattal,~D.~A.; Gauger,~N.~R.; Fan,~S.; Beausoleil,~R.~G. Sensitivity analysis and optimization of sub-wavelength optical gratings using adjoints. \emph{Optics express} \textbf{2014}, \emph{22}, 12971--12981\relax
\mciteBstWouldAddEndPuncttrue
\mciteSetBstMidEndSepPunct{\mcitedefaultmidpunct}
{\mcitedefaultendpunct}{\mcitedefaultseppunct}\relax
\EndOfBibitem
\bibitem[Svanberg(2002)]{Svanberg2002}
Svanberg,~K. A class of globally convergent optimization methods based on conservative convex separable approximations. \emph{SIAM journal on optimization} \textbf{2002}, \emph{12}, 555--573\relax
\mciteBstWouldAddEndPuncttrue
\mciteSetBstMidEndSepPunct{\mcitedefaultmidpunct}
{\mcitedefaultendpunct}{\mcitedefaultseppunct}\relax
\EndOfBibitem
\bibitem[nlo()]{nlopt}
The {NLopt} nonlinear-optimization package. \url{http://github.com/stevengj/nlopt}\relax
\mciteBstWouldAddEndPuncttrue
\mciteSetBstMidEndSepPunct{\mcitedefaultmidpunct}
{\mcitedefaultendpunct}{\mcitedefaultseppunct}\relax
\EndOfBibitem
\bibitem[Zhou \latin{et~al.}(2015)Zhou, Lazarov, Wang, and Sigmund]{Zhou15}
Zhou,~M.; Lazarov,~B.; Wang,~F.; Sigmund,~O. Minimum length scale in topology optimization by geometric constraints. \emph{Computer Methods in Applied Mechanics and Engineering} \textbf{2015}, \emph{293}, 266--282\relax
\mciteBstWouldAddEndPuncttrue
\mciteSetBstMidEndSepPunct{\mcitedefaultmidpunct}
{\mcitedefaultendpunct}{\mcitedefaultseppunct}\relax
\EndOfBibitem
\bibitem[Li \latin{et~al.}(2016)Li, Chen, Liu, and Tong]{Li2016}
Li,~Q.; Chen,~W.; Liu,~S.; Tong,~L. Structural topology optimization considering connectivity constraint. \emph{Structural and Multidisciplinary Optimization} \textbf{2016}, \emph{54}\relax
\mciteBstWouldAddEndPuncttrue
\mciteSetBstMidEndSepPunct{\mcitedefaultmidpunct}
{\mcitedefaultendpunct}{\mcitedefaultseppunct}\relax
\EndOfBibitem
\bibitem[Jensen and Sigmund(2005)Jensen, and Sigmund]{jensen2005damping}
Jensen,~J.~S.; Sigmund,~O. Topology optimization of photonic crystal structures: a high-bandwidth low-loss T-junction waveguide. \emph{JOSA B} \textbf{2005}, \emph{22}, 1191--1198\relax
\mciteBstWouldAddEndPuncttrue
\mciteSetBstMidEndSepPunct{\mcitedefaultmidpunct}
{\mcitedefaultendpunct}{\mcitedefaultseppunct}\relax
\EndOfBibitem
\bibitem[Hansen \latin{et~al.}(2024)Hansen, Arregui, Babar, Christiansen, and Stobbe]{hansen2024inverse}
Hansen,~S.~E.; Arregui,~G.; Babar,~A.~N.; Christiansen,~R.~E.; Stobbe,~S. Inverse design and characterization of compact, broadband, and low-loss chip-scale photonic power splitters. \emph{Materials for Quantum Technology} \textbf{2024}, \emph{4}, 016201\relax
\mciteBstWouldAddEndPuncttrue
\mciteSetBstMidEndSepPunct{\mcitedefaultmidpunct}
{\mcitedefaultendpunct}{\mcitedefaultseppunct}\relax
\EndOfBibitem
\bibitem[Blondel \latin{et~al.}(2022)Blondel, Berthet, Cuturi, Frostig, Hoyer, Llinares-L{\'o}pez, Pedregosa, and Vert]{blondel2022efficient}
Blondel,~M.; Berthet,~Q.; Cuturi,~M.; Frostig,~R.; Hoyer,~S.; Llinares-L{\'o}pez,~F.; Pedregosa,~F.; Vert,~J.-P. Efficient and modular implicit differentiation. \emph{Advances in neural information processing systems} \textbf{2022}, \emph{35}, 5230--5242\relax
\mciteBstWouldAddEndPuncttrue
\mciteSetBstMidEndSepPunct{\mcitedefaultmidpunct}
{\mcitedefaultendpunct}{\mcitedefaultseppunct}\relax
\EndOfBibitem
\bibitem[Boyd and Vandenberghe(2004)Boyd, and Vandenberghe]{boyd2004convex}
Boyd,~S.; Vandenberghe,~L. \emph{Convex Optimization}; Cambridge University Press, 2004\relax
\mciteBstWouldAddEndPuncttrue
\mciteSetBstMidEndSepPunct{\mcitedefaultmidpunct}
{\mcitedefaultendpunct}{\mcitedefaultseppunct}\relax
\EndOfBibitem
\end{mcitethebibliography}
\end{document}